\documentclass[conference]{IEEEtran}
\IEEEoverridecommandlockouts
\usepackage{algorithm}
\usepackage{algpseudocode}
\algrenewcommand\algorithmiccomment[1]{\hspace*{\fill}\(\triangleright\)\, #1}

\usepackage{tabularx}
\usepackage{array}
\usepackage{cite}
\usepackage{amsmath,amssymb,amsfonts}
\usepackage{multirow}
\usepackage{graphicx}
\usepackage{textcomp}
\usepackage{xcolor}
\usepackage{booktabs}

\def\BibTeX{{\rm B\kern-.05em{\sc i\kern-.025em b}\kern-.08em
    T\kern-.1667em\lower.7ex\hbox{E}\kern-.125emX}}
\begin{document}

\title{Learning to Sync at the Edge: Client-Side Adaptive Synchronization for Federated Learning}

\author{
\IEEEauthorblockN{
Ebasa Temesgen\textsuperscript{},
Ammar Ahmed\textsuperscript{},
Nathan Lechamo\textsuperscript{\dag},
Ali Anwar\textsuperscript{},
Maria Gini\textsuperscript{}
}
\IEEEauthorblockA{
\textsuperscript{}Department of Computer Science \& Engineering,
University of Minnesota, Minneapolis, MN, USA\\
\textsuperscript{\dag}Department of Computer Science,
University of Texas at San Antonio, San Antonio, TX, USA\\
\{temes021, ahme0599, aanwar, gini\}@umn.edu,
nathan.lechamo@utsa.edu
}
}

\maketitle

\begin{abstract}
The scalability and performance of edge federated learning are fundamentally constrained by synchronization bottlenecks. Clients operate under heterogeneous compute, time-varying communication conditions, and, in some settings, privacy constraints, yet synchronization is often still handled with fixed schedules. However, synchronization decisions are often made outside the client, even though the client directly observes local progress, resource state, and communication conditions. We present a client-side reinforcement learning synchronization framework in which each client decides during training whether to upload, download, synchronize bidirectionally, or defer communication. The framework places a lightweight scheduler alongside the local model and formulates synchronization as a resource-aware control problem driven by local learning and system state. We study this framework on two representative edge workloads, spatiotemporal traffic forecasting and image classification. Across both settings, adaptive client-side synchronization improves the accuracy--cost trade-off, reducing system cost in traffic forecasting while maintaining competitive cost and improving final accuracy in image classification. We also consider privacy-constrained settings, where synchronization must account for not only latency and energy but also privacy expenditure. Overall, our results show that synchronization in edge federated learning is better treated as an adaptive client-side control problem than only as a server-side coordination mechanism.
\end{abstract}


\section{Introduction}

Federated learning (FL) is appealing for edge computing because it keeps raw data on-device while allowing many resource-constrained clients to train a shared model collaboratively~\cite{mcmahan2017communication}. This is especially valuable in edge settings such as traffic sensing, mobility analytics, and distributed perception, where data are generated continuously across many devices and may be costly, sensitive, or impractical to move to a central location. By keeping data local, FL reduces repeated raw-data transfer while still enabling clients to benefit from shared learning.

In practice, however, the limiting factor is often not local optimization itself, but synchronization. Edge clients rarely operate under stable conditions: wireless bandwidth drifts over time, devices differ in compute capability and energy budget, and local data streams evolve. In privacy-sensitive settings such as traffic and mobility data, communication can also expose sensitive model updates and metadata~\cite{bonawitz2019towards,de2013unique}. Despite this, synchronization control is often handled at the server or as part of the training protocol, even though many of the signals that determine whether communication is useful are local to the client. A client may have made little local progress, may be operating under a degraded channel, or may be close to exhausting a privacy budget. Synchronizing in these states can add latency, energy use, and privacy exposure without proportional learning benefit.

\begin{figure}[t]
    \centering
\includegraphics[width=\linewidth]{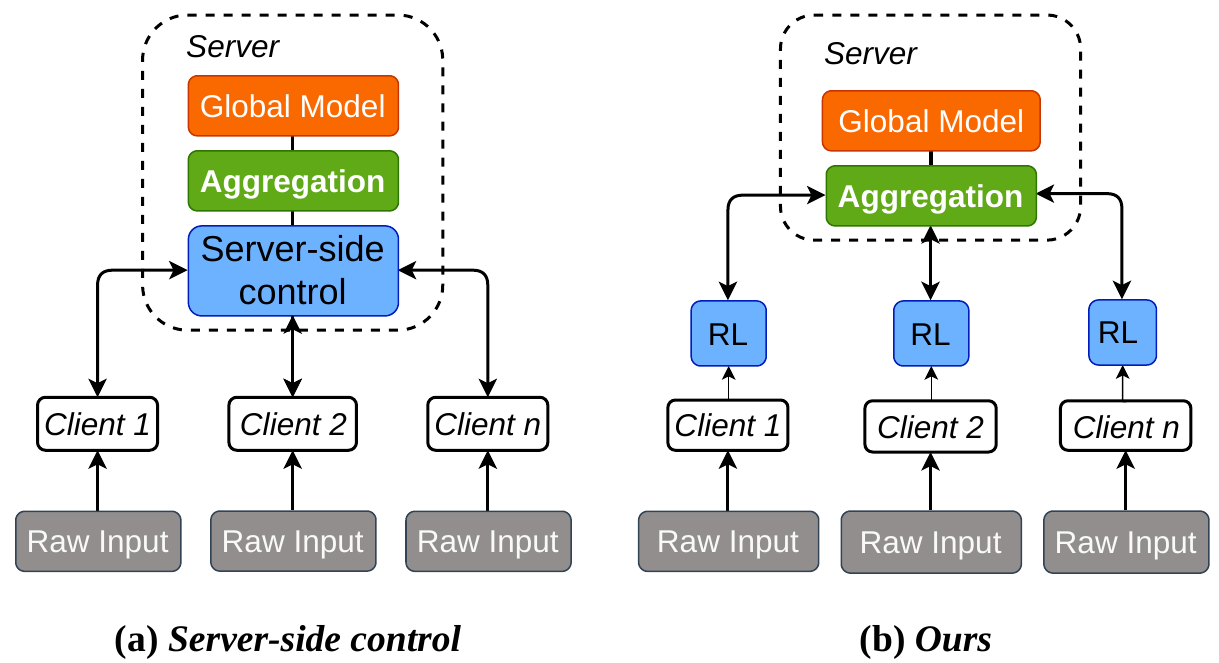}
    \caption{Server-side versus client-side communication control in federated learning. (a) Prior work typically adapts communication from the server. (b) Our approach places a lightweight scheduler at each client to make synchronization decisions from local state.}
    \label{fig:arch}
\vspace{-1em}
\end{figure}

This paper studies a central systems question in edge federated learning: \textit{when should a client synchronize}? Rather than treating synchronization as only a server-side coordination decision, we treat it as a decision-event made at the client. Prior approaches primarily adapt communication from the server, for example through aggregation timing, buffering, or client selection~\cite{xie2019asynchronous,nguyen2022federated,lai2021oort}. In contrast, our framework (shown in Figure~\ref{fig:arch}) moves synchronization control to the client, where the decision can respond directly to local utility and resources. 

Learning-based client selection has also been studied using
multi-agent PPO for dynamic client selection in vehicular
federated learning~\cite{yu2024multi}. Our focus is the synchronization action available to a client: skip communication, upload, download, or synchronize bidirectionally.  This motivates viewing synchronization as a client-side control
problem rather than only a server-side scheduling or aggregation problem. The value of synchronizing depends jointly on local learning progress, staleness, channel state, energy cost, privacy budget, and the future consequences of acting now versus waiting. A decision that saves cost in the current round may increase staleness and reduce utility later, while immediate synchronization may improve coordination at the expense of time and energy. These delayed and coupled trade-offs are difficult to capture with a single hand-designed rule or fixed protocol, since simple threshold policies typically react to only one signal, such as model change, staleness, or bandwidth.

We therefore introduce a lightweight reinforcement learning (RL) scheduler for this client-side decision process, as illustrated in Figure~\ref{fig:arch}, and evaluate it against learning-based client schedulers and server-side communication-control schedulers. At each decision event, the scheduler observes client-side signals that reflect both learning status and system conditions, including indicators of model utility, communication cost, and client resource state. Using compact server-broadcast context such as model update freshness, rank, and recent synchronization rates further improves performance, while the action remains selected by the client scheduler. The scheduler then selects whether to upload, download, synchronize bi-directionally, or defer communication. This allows synchronization to respond to learning progress, device heterogeneity, and time-varying network conditions without changing the underlying federated training pipeline. In this way, communication becomes an adaptive systems-level mechanism rather than a static training parameter.

We also consider settings in which synchronization cost is shaped not only by latency and energy, but also by privacy constraints. This motivates a privacy-aware formulation of client-side synchronization, in which communication decisions are made with awareness of both system cost and privacy considerations rather than treating privacy as a separate concern after the schedule has already been chosen.

We examine our framework on two representative edge workloads: spatiotemporal traffic forecasting and image classification. These workloads differ substantially in data characteristics, learning dynamics, and operating conditions, allowing us to evaluate whether the same client-side synchronization mechanism can perform across distinct edge learning settings. This paper makes the following contributions:
\begin{itemize}
    \item We present a client-side adaptive synchronization RL framework for federated learning at the edge, in which each client decides when to synchronize with the server.
    
    \item We formulate synchronization as a resource-aware control problem that combines task utility with communication latency, energy cost, device heterogeneity, and privacy constraints.
    
    \item We evaluate the same framework on two representative edge workloads, traffic forecasting and image classification, against learning-based and server-side communication-control baselines.
    
    \item We show the scheduler improves the accuracy--cost trade-off across traffic forecasting and image classification.
\end{itemize}

\section{Related Work}

Communication efficiency remains a central challenge in federated learning,
especially under bandwidth constraints, heterogeneous devices, intermittent
availability, and non-IID data. Prior work has addressed this challenge
through several control points. Server-side asynchronous methods such as
FedAsync~\cite{xie2019asynchronous} and
FedBuff~\cite{nguyen2022federated} control when received updates are applied
or buffered. Client-selection methods such as Oort~\cite{lai2021oort} and
RL-based schedulers such as FedMarl~\cite{zhang2022multi} decide which
clients should participate. Optimization methods such as
SCAFFOLD~\cite{karimireddy2020scaffold} and
FedProx~\cite{li2020federated} reduce the effect of statistical
heterogeneity during local training. Other work relaxes rigid round-based
execution through arbitrary communication patterns and flexible device
participation~\cite{avdiukhin2021federated,ruan2021towards}. These methods
improve federated learning under communication and heterogeneity constraints,
but they primarily control aggregation, participation, protocol flexibility,
or local optimization. In contrast, our work controls the synchronization action taken by an available client by deciding whether to skip communication, upload a local update, download the global model, or synchronize bidirectionally.

\begin{table}[t]
\centering
\caption{Communication-control mechanisms in federated learning.}
\label{tab:comm_control}
\footnotesize
\setlength{\tabcolsep}{3pt}
\renewcommand{\arraystretch}{1.08}
\begin{tabular}{p{2.35cm} p{1.95cm} p{1.85cm} p{2cm}}
\toprule
\textbf{Method} & \textbf{Control point} & \textbf{Controller} & \textbf{Decision} \\
\midrule
FedAsync & Aggregation & Server & Apply updates \\
FedBuff & Aggregation & Server & Buffer updates \\
Oort & Participation & Server & Select clients \\
FedMarl & Participation & Server & Schedule clients \\
SCAFFOLD & Optimization & Server/client & Correct drift \\
FedProx & Optimization & Client-objective & Stabilize training \\
Avdiukhin et al.~\cite{avdiukhin2021federated} & Protocol & Server/system & Arbitrary comm. \\
Ruan et al.~\cite{ruan2021towards} & Participation & Server/system & Availability \\
\textbf{Ours} & \textbf{Synchronization} & \textbf{Client} & \textbf{Sync action} \\
\bottomrule
\end{tabular}
\end{table}

Table~\ref{tab:comm_control} summarizes this distinction. Our scheduler is
client-side, but the main distinction is the synchronization action space:
the policy separates upload, download, bidirectional synchronization, and
skipping. This lets each client react to local learning progress, model
staleness, communication quality, energy cost, and privacy budget without
changing the server aggregation rule.

Federated learning is attractive for edge systems because it allows many
distributed clients to learn collaboratively without moving raw data to a
central server. However, keeping data local does not eliminate the cost of
synchronization. Model updates still consume bandwidth and energy, and may
require additional privacy protection. Prior FL systems~\cite{bonawitz2019towards,geyer2017differentially}
study secure aggregation and differentially private updates, while
resource-aware methods such as FedCS~\cite{nishio2019client} account for
client compute and wireless conditions. Our work uses these constraints at a
different control point: we expose communication cost, energy cost,
staleness, and privacy budget as local decision signals for client-side
synchronization.

These issues are especially relevant in traffic forecasting, where sensors
generate continuous, spatially coupled streams under heterogeneous local
conditions. Prior federated traffic systems include
FCGCN~\cite{xia2022short}, FedAGCN~\cite{qi2023fedagcn}, and
FedTFP~\cite{feng2024traffic}. These works focus primarily on federated
forecasting quality, whereas our focus is the client-side synchronization
decision under changing bandwidth, energy, staleness, and privacy costs.

\section{System Overview and Design}
\label{sec:system_overview}

\begin{figure*}[t]
    \centering
    \includegraphics[width=0.95\linewidth]{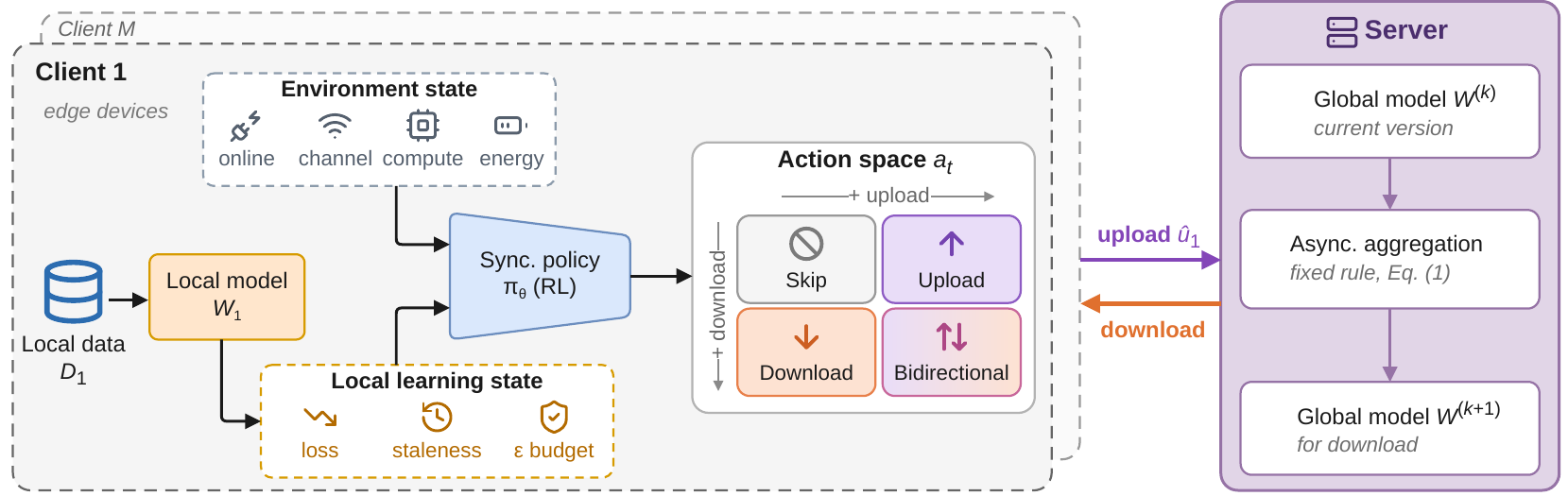}
    \caption{System overview of the proposed client-side synchronization scheduler. Each client trains a local task model on private data and uses an RL policy to decide whether to skip, upload, download, or synchronize bidirectionally based on local learning state and environment conditions. The server maintains the global model and aggregates received client updates.
  }
    \label{fig:system-architecture}
\end{figure*}

We consider an edge federated learning system with a central server and
heterogeneous clients. Each client stores private data, trains a local task
model, and runs a lightweight synchronization scheduler. The server maintains
the global model and aggregates received updates, while the scheduler decides
when communication is worth its latency, energy, and privacy cost.

Our design separates task learning from synchronization control. The task
model performs prediction or classification, while the scheduler controls
only communication with the server. This separation lets the same scheduling
mechanism support different task backbones. We use STAEFormer~\cite{liu2023spatio}
for traffic forecasting and ResNet-18~\cite{He2016} for image classification, but the
scheduler itself is model-agnostic.

Figure~\ref{fig:system-architecture} shows the architecture. At each decision
event, the scheduler observes local learning and system state and chooses
whether to upload, download, synchronize bidirectionally, or defer
communication. For our method, the server update rule is fixed so that improvements come
from client-side synchronization rather than changing the aggregation rule.

\subsection{Event-Driven Workflow}
The system is simulated as an event-driven process. Each client evolves
through five runtime states: unavailable, available-idle, local-training,
uploading, and downloading. A priority queue of timestamped events advances the simulation in event time, including availability changes, local-training completion, scheduler decisions, communication completion, and server updates.

Local training is scheduled automatically when a client is available, idle, and has no pending communication. Scheduler decisions are generated only for available idle clients, which prevents the policy from interrupting training or in-flight communication. Thus, the scheduler controls synchronization
rather than local training.

At a decision event, client \(m\) maintains a local model \(W_m\) trained on private data \(D_m\). Upload transmits a protected model update, download refreshes the local model from the server, and the bidirectional action uploads, waits for the corresponding server update, and then downloads the updated global model.

\subsection{Server Aggregation}
For our client-side scheduler, we keep the server update rule fixed so that
performance differences come from the synchronization policy rather than an aggregation algorithm.

The server maintains a global model \(W^{(k)}\), where \(k\) indexes
server update events. Because clients operate asynchronously, an
update may be based on an older server version. Let \(b_{m,k}\) be
the server version last downloaded by client \(m\) before producing
the update available at event \(k\). We define
\[
u_{m,k}=W_{m,k}^{\mathrm{loc}}-W^{(b_{m,k})},
\qquad
s_{m,k}=k-b_{m,k},
\]
where \(W_{m,k}^{\mathrm{loc}}\) is the client model after local
training and \(s_{m,k}\) is the update staleness.

Let \(\hat u_{m,k}\) be the protected update after optional clipping
and Gaussian perturbation. At server event \(k\), let
\(\mathcal{M}_{\mathrm{up}}^{(k)}\) denote the set of client uploads
applied at that event. The server step size is \(\eta_s\), and
\(\alpha(s_{m,k})\) denotes an optional staleness weight. The server
applies
\begin{equation}
W^{(k+1)}
=
W^{(k)}
+
\eta_s
\sum_{m \in \mathcal{M}_{\mathrm{up}}^{(k)}}
\omega_{m,k}
\alpha(s_{m,k})
\hat u_{m,k},
\label{eq:fedavg_overview}
\end{equation}
where the data-size weight is
\[
\omega_{m,k}
=
\frac{|\mathcal{D}_m|}
{\sum_{j \in \mathcal{M}_{\mathrm{up}}^{(k)}} |\mathcal{D}_j|}.
\]
If privacy is disabled, \(\hat u_{m,k}=u_{m,k}\). The server update rule is fixed within each configuration; the image experiments use \(\eta_s=0.2\). Thus, uploaded updates are
applied directly, and staleness is handled by the client-side scheduler
through its observation and reward rather than by a server-side
staleness correction rule.

\subsection{Cost, Dynamics, and Privacy}

The scheduler evaluates communication against the conditions that a client
experiences at the decision time. Following standard wireless cost
models~\cite{wang2018federated,wu2021load}, an update of \(\Delta_m\) bits sent
at rate \(R_m^{\mathrm{up}}\) incurs
\begin{equation}
T_m^{\mathrm{up}}=\frac{\Delta_m}{R_m^{\mathrm{up}}},\qquad
E_m^{\mathrm{up}}=P_mT_m^{\mathrm{up}},
\label{eq:compact_uplink}
\end{equation}
with analogous download costs. Effective rates vary with good, medium, and
degraded channel regimes.

\begin{figure}[t]
\centering
\includegraphics[width=\linewidth]{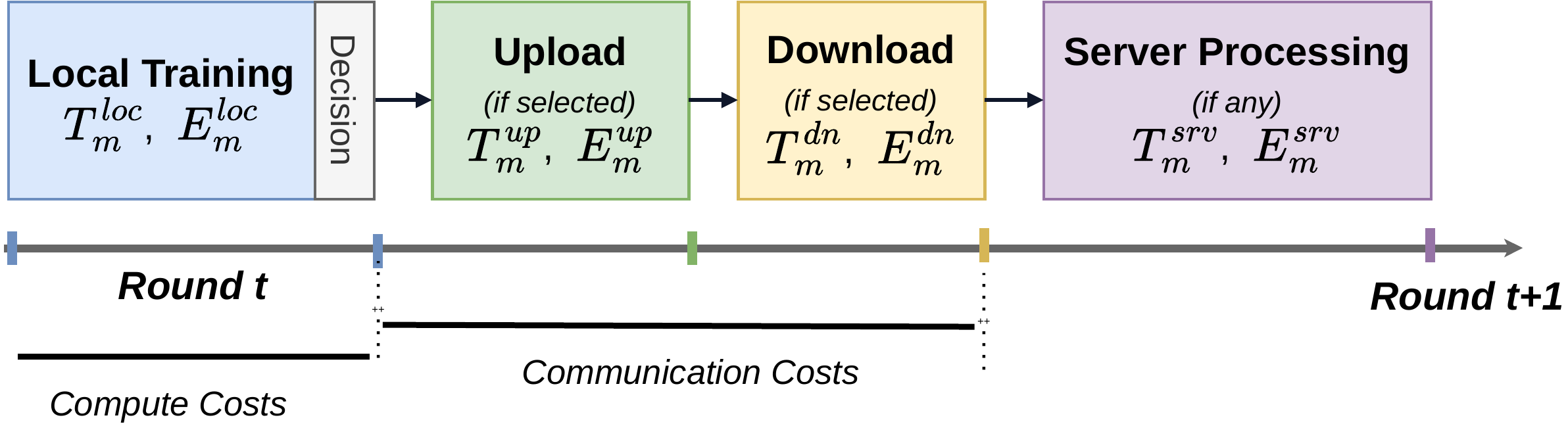}
\caption{Cost accounting at one client's decision.}
\label{fig:compact_timeline}
\end{figure}

Client computation is charged separately from communication. For a CPU client
that processes \(n_m(t)\) samples at frequency \(f_m\), with \(q\) cycles per
sample and hardware coefficient \(\kappa\), we use
\begin{equation}
T_m^{\rm loc}(t)=\frac{qn_m(t)}{f_m},\qquad
E_m^{\rm loc}(t)=\kappa f_m^2qn_m(t).
\label{eq:compact_cpu}
\end{equation}
For a GPU profile, a Roofline-style bound~\cite{williams2009roofline} accounts
for both arithmetic throughput \(\theta_m\) and memory bandwidth \(\beta_m\):
\begin{equation}
T_{m,\rm gpu}^{\rm loc}=\max\!\left(\frac{F_m}{\theta_m},
\frac{M_m}{\beta_m}\right),\qquad
E_{m,\rm gpu}^{\rm loc}=P_m^{\rm avg}T_{m,\rm gpu}^{\rm loc}.
\label{eq:compact_gpu}
\end{equation}
This separation prevents communication savings from being conflated with
faster client hardware. The event engine also models stable, cyclic, and
bursty availability. Availability profiles and channel regimes jointly
determine whether an action is feasible and how long it takes; future states
are not revealed to the policy. Figure~\ref{fig:compact_timeline} summarizes the per-round sequence.

The image experiments clip each upload to \(C\). The server averages
\(B_{\rm agg}\) buffered uploads and adds Gaussian noise before release:
\begin{align}
\widetilde u_j&=\frac{u_j}{\max(1,\lVert u_j\rVert_2/C)},\notag\\
\widehat u_k&=\frac{1}{B_{\rm agg}}\sum_{j\in\mathcal B_k}\widetilde u_j
+\mathcal N\!\left(0,\frac{\sigma^2 C^2}{B_{\rm agg}^2}I\right).
\label{eq:compact_dp}
\end{align}
Here \(\mathcal B_k\) indexes buffered uploads, including repeated client
contributions. 
Gaussian RDP accounting~\cite{mironov2017renyi} uses \(\delta=10^{-5}\),
composes noisy releases across episodes without subsampling amplification,
and broadcasts spent/remaining budget to clients. Uploads exceeding
\(\varepsilon_{\max}=12\) are masked. Reported \(\varepsilon\) values
characterize this release accountant.

\section{Client-Side Synchronization Policy}
\label{sec:compact_policy}

\begin{figure}[h]
    \centering
\includegraphics[width=0.9\linewidth]{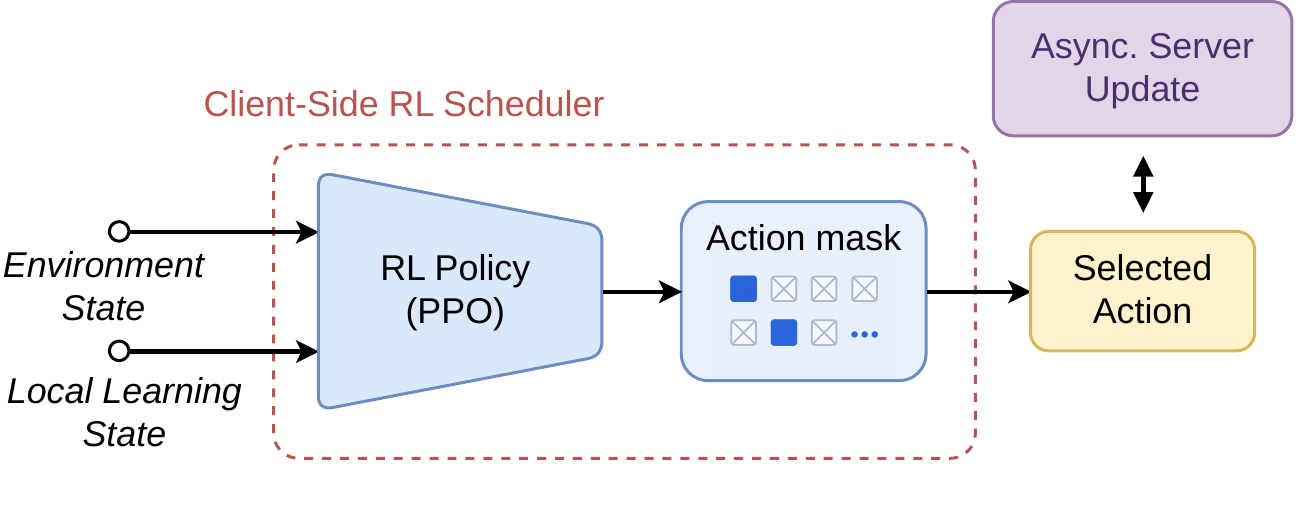}
    \caption{Observation and action-selection pipeline for the client-side RL scheduler.}
    \label{fig:rl_environment_flow}
\end{figure}

At decision event \(t\), client \(m\) selects
\(a_t^{(m)}\in\{\mathsf S,\mathsf U,\mathsf D,\mathsf B\}\), denoting skip,
upload, download, and bidirectional synchronization. All clients share one
small policy, but evaluate it on their own observations. The observation
contains four groups: (i) current availability, channel regime, and estimated
transfer cost; (ii) local loss, loss change, and compute cost; (iii) dirty
state, upload/download age, and server--client version gap; and (iv) spent and
remaining privacy budget. The accuracy-oriented policy may additionally use
compact server broadcasts for freshness, buffer state, client rank, and recent
synchronization rates, but action selection remains on the client. At each decision event, the scheduler receives a client-level observation composed of environment state and local client state, as shown in Figure~\ref{fig:rl_environment_flow}. Table~\ref{tab:compact_observation} lists the observation groups. The communication group represents environment state, capturing availability and channel-dependent communication cost. The learning, synchronization, and privacy groups represent local client state, capturing model progress, synchronization history, and privacy budget.

\begin{table}[h]
\caption{Client observation supplied to the shared policy.}
\label{tab:compact_observation}
\centering
\scriptsize
\setlength{\tabcolsep}{3pt}
\renewcommand{\arraystretch}{1.08}
\begin{tabular}{@{}p{1.45cm}p{5.75cm}@{}}
\toprule
Group & Features\\
\midrule
Comm. & Availability, channel regime, transfer time and energy\\
Learning & Loss, loss change, dirty flag, latest compute cost\\
Sync. & Upload/download age, local/server versions, staleness\\
Privacy & Spent/remaining budget and most recent increment\\
\bottomrule
\end{tabular}
\end{table}

Legal-action masking encodes the runtime constraints. Skip is always legal;
upload requires uncommitted local progress, a satisfied minimum communication
gap, and sufficient privacy budget; download requires a newer server version;
and bidirectional synchronization requires a legal upload. The mask is applied
both when sampling actions and when recomputing the learning loss. Consequently,
the policy never learns from actions that the event engine could not execute.

\subsection{Staleness and Uncommitted Progress}

A useful communication decision must distinguish an old model from a model
that contains new local information. Let \(\tau_m(t)\) be the number of client
decision events since the most recent upload, and let \(v_t^{\rm srv}\) and
\(v_t^{\rm loc,m}\) be the current server version and the newest server version
incorporated by client \(m\). Upload age and the server--client version gap are
\begin{align}
\tau_m(t+1)&=
\begin{cases}
0,&I_{\rm up,t}^{(m)}=1,\\
\tau_m(t)+1,&I_{\rm up,t}^{(m)}=0,
\end{cases}\notag\\
\nu_t^{(m)}&=\max(v_t^{\rm srv}-v_t^{\rm loc,m},0).
\label{eq:compact_staleness}
\end{align}
The first term measures how long the server has waited for this client; the
second measures how far the client's reference model trails the server. Staleness alone cannot reveal whether waiting withholds useful work. If
\(t_m^{\rm last}\) is the last upload event, relative uncommitted progress is
\begin{equation}
p_t^{(m)}=\frac{\lVert W_m^{(t)}-W_m^{(t_m^{\rm last})}\rVert_2}
 {\lVert W_m^{(t_m^{\rm last})}\rVert_2+\eta},\qquad \eta>0.
\label{eq:compact_dirty_progress}
\end{equation}
We therefore use
\begin{align}
c_{\rm stale,t}^{(m)}&=\alpha_\tau\tau_m(t)+\alpha_\nu\nu_t^{(m)},\notag\\
c_{\rm dirty,t}^{(m)}&=(1-I_{\rm up,t}^{(m)})p_t^{(m)}(1+\tau_m(t)).
\label{eq:compact_structured_costs}
\end{align}


Skipping remains inexpensive when little has changed, but becomes costly when
fresh local progress is repeatedly withheld. An optional channel term
\(c_{\rm chan}=I_{\rm up}[R_{\min}^{\rm up}/R^{\rm up}-1]_+\) discourages an
avoidable upload over a degraded link. In private runs, an upload also incurs
the accountant's increment \(\varepsilon_t^{(m)}\), subject to
\(\sum_t I_{\rm up,t}^{(m)}\varepsilon_t^{(m)}\leq
\varepsilon_m^{\max}\). These structured terms explain why the policy cannot
optimize the task loss or byte count in isolation.

The reward unifies task and systems objectives. With upload/download indicators
\(I_{\mathrm{up}}\) and \(I_{\mathrm{dn}}\), we define
\begin{align}
\mathcal C_t^{(m)}={}&\lambda_\ell c_\ell
+\lambda_T(T^{\mathrm{loc}}+I_{\mathrm{up}}T^{\mathrm{up}}
 +I_{\mathrm{dn}}T^{\mathrm{dn}}) \notag\\
&+\lambda_E(E^{\mathrm{loc}}+I_{\mathrm{up}}E^{\mathrm{up}}
 +I_{\mathrm{dn}}E^{\mathrm{dn}}) \notag\\
&+\lambda_s c_{\mathrm{stale}}+\lambda_d c_{\mathrm{dirty}}
+\lambda_p I_{\mathrm{up}}\varepsilon_t+\lambda_c c_{\mathrm{chan}},
\label{eq:compact_cost}
\end{align}
where fixed reference scales normalize the terms. The dirty-state penalty grows
when useful local progress is repeatedly withheld; the channel penalty
discourages uploads below a target rate. The reward is
\(r_t^{(m)}=-\mathcal C_t^{(m)}\), making the control objective consistent with
end-to-end resource cost rather than communication count alone.

We optimize the shared actor and value function with masked Proximal Policy
Optimization (PPO)~\cite{schulman2017proximal}. For advantage estimate \(A_t\),
the actor maximizes
\begin{align}
L^{\mathrm{clip}}(\theta)
&=\mathbb E_t\!\left[\min\!\left(\rho_tA_t,
\operatorname{clip}(\rho_t,1-\epsilon,1+\epsilon)A_t\right)\right],\notag\\
\rho_t&=\frac{\pi_\theta(a_t\mid o_t)}
 {\pi_{\mathrm{old}}(a_t\mid o_t)}.
\label{eq:compact_ppo}
\end{align}
Generalized advantage estimation \cite{schulman2015high} assigns credit across later communication and
learning events. The deployed actor has 20.4k parameters (0.081 MB in FP32), so
its size is independent of the number of clients.

For a finite rollout, the critic supplies the bootstrap needed to evaluate
actions whose learning benefit arrives after the current communication event.
With temporal-difference residual
\(\delta_t=r_t+\gamma V_\phi(o_{t+1})-V_\phi(o_t)\), truncated generalized
advantage estimation computes
\begin{equation}
A_t=\sum_{j=0}^{L_t-1}(\gamma\lambda_{\rm GAE})^j\delta_{t+j}.
\label{eq:compact_gae}
\end{equation}
The value target is the observed discounted reward plus the final bootstrap,
\begin{equation}
G_t=\sum_{j=0}^{L_t-1}\gamma^jr_{t+j}
+\gamma^{L_t}V_\phi(o_{t+L_t}),
\label{eq:compact_return}
\end{equation}
and the critic minimizes \(\mathbb E_t[(V_\phi(o_t)-G_t)^2]\). Keeping actor
and critic targets tied to event order is important: a costly upload can become
valuable only after its server update affects later clients.

Algorithm~\ref{alg:compact_ppo} shows how event simulation and policy learning
interleave. A rollout contains decisions from whichever client becomes ready
next, rather than a synchronous vector of one action per round. The value
function is trained against bootstrapped returns, while the actor reuses the
stored legal-action mask during every PPO epoch. This preserves the behavior
policy's support and prevents invalid actions from re-entering through loss
recomputation.

\begin{algorithm}[t]
\caption{Event-driven client-side PPO}
\label{alg:compact_ppo}
\footnotesize
\begin{algorithmic}[1]
\State Initialize server \(W\), clients, event queue, actor \(\pi_\theta\), critic \(V_\phi\)
\For{each PPO iteration}
  \State rollout \(\Gamma\gets\emptyset\)
  \While{\(|\Gamma|<T_s\)}
    \State \((m,o,M)\gets\textsc{NextReadyClient}()\)
    \State sample \(a\sim\pi_\theta(\cdot\mid o,M)\)
    \State \((o',\mathcal C)\gets\textsc{AdvanceEvents}(m,a)\)
    \State append \((o,a,-\mathcal C,o',M)\) to \(\Gamma\)
  \EndWhile
  \State compute bootstrapped returns and GAE advantages
  \State update \(\theta\) with Eq.~\eqref{eq:compact_ppo}; fit \(V_\phi\)
\EndFor
\end{algorithmic}
\end{algorithm}

\section{Evaluation}
\label{sec:compact_evaluation}

\begin{figure}[t]
    \centering
    \includegraphics[width=.92\linewidth]{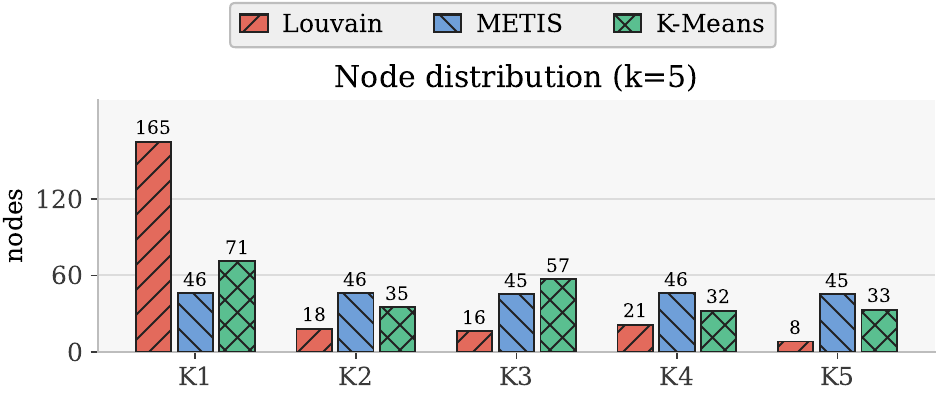}
    \caption{Client-size balance for five-way PeMSD7(M) partitioning
(228 sensors)~\cite{yu2018stgcn}.}
    \label{fig:metis_sizes}
\end{figure}

\textbf{Workloads.}
The traffic workflow uses PeMS-08 measurements sampled every five
minutes~\cite{Songetal2020}.
For client formation, we partition the traffic sensor graph using METIS~\cite{karypis1997metis} as it represents the spatial elements of our data. We compare METIS with two common alternatives, K-Means~\cite{HartiganandWong1979} and Louvain~\cite{Blondeletal2008} community detection. In our setup, METIS produces more balanced clusters while preserving the graph structure of the road network. This is useful for federated traffic forecasting because extreme client-size imbalance can affect aggregation behavior, and communication load. Figure~\ref{fig:metis_sizes} shows the resulting partition-size comparison.

We train federated STAEFormer (FedSTAE) on four clients, and evaluate
200 episodes. The image workflow uses a CIFAR-10 \cite{krizhevsky2009learning} ResNet-18 with a
\(3\times3\) input convolution, no initial max-pooling, and a ten-class head.
20 clients receive a Dirichlet non-IID split with \(\alpha=0.3\), and each
run lasts 300 episodes. These workloads exercise the same action interface and
policy architecture while changing the task model and utility signal.

\textbf{Comparisons and protocol.}
We compare with FedAsync, FedBuff, and Oort as representative server-side
controls; DQN~\cite{Mnihetal2015} as a client scheduler with the same observation,
action space, and legal-action mask; and Always-sync as a fixed client-side
policy. Each baseline uses its validation-selected configuration on the same
data split. Image results are means across completed runs; evaluation frequency and
test-set coverage vary across runs. Traffic results report a
validation-selected completed run, with an average over its evaluation
horizon. Experiments use an RTX 4090 and 24 CPU threads.

Table~\ref{tab:compact_hyperparameters} reports the default scheduler and
environment configuration. We report
MAE/RMSE for traffic, accuracy for images, and latency, energy, bytes, and
synchronization counts for system cost.

\begin{table}[t]
\caption{Scheduler and environment configuration.}
\label{tab:compact_hyperparameters}
\centering
\scriptsize
\setlength{\tabcolsep}{3pt}
\renewcommand{\arraystretch}{1.06}
\begin{tabular}{@{}p{2.0cm}p{5.15cm}@{}}
\toprule
Setting & Value\\
\midrule
PPO objective & $\gamma=0.99$, GAE $\lambda=0.95$, clip $0.2$\\
PPO optimizer & LR $3\!\times\!10^{-4}$, 6 epochs, batch 64,
entropy $0.006$, value coefficient $0.5$\\
Channel & $B=10^5$ Hz; multipliers $1.5/1.0/0.45$ with
probabilities $0.35/0.45/0.20$\\
Device & $P_{\rm tx}=1.5$ W, $P_{\rm rx}=0.8$ W;
$f_m\sim U(0.6,1.0)$ GHz\\
Privacy & DP; $C=1$, $\sigma=6$,
$\delta=10^{-5}$, $\varepsilon_{\max}=12$\\
Task weights & $\lambda_{\rm loss}=0.35$,
$\lambda_{\rm time}=0.75$, $\lambda_{\rm ene}=1.0$\\
System weights & $\lambda_{\rm stale}=0.08$,
$\lambda_{\rm dirty}=0.18$, $\lambda_{\rm chan}=0.25$,
$\lambda_{\rm priv}=0.02$\\
Local optimizer & SGD, LR $0.02$, momentum $0.9$, weight decay
$5\!\times\!10^{-4}$, gradient clip $5.0$\\
\bottomrule
\end{tabular}
\end{table}

\section{Results}
\label{sec:compact_results}

\begin{table*}[t]
\caption{Traffic forecasting on PeMS-08 with four clients over 200 episodes.
The validation-selected run is averaged over the evaluation horizon; lower is
better for errors and costs.}
\label{tab:compact_traffic}
\centering
\scriptsize
\setlength{\tabcolsep}{4pt}

\begin{tabular*}{0.9\textwidth}{@{\extracolsep{\fill}}lrrrrrrrrr@{}}
\toprule
Method & MAE & RMSE & MAPE & Latency (s) & Energy (J) & Uploaded (GB) & Uploads & Downloads & Agg.\\
\midrule
DQN      &20.44&30.78&\textbf{14.85}&$1.4\!\times\!10^5$&$1.8\!\times\!10^5$&28.37&6,439&3,552&6,226\\
FedAsync &20.51&31.34&15.33&$1.4\!\times\!10^5$&$1.9\!\times\!10^5$&30.56&6,936&3,657&6,712\\
FedBuff  &22.12&32.67&18.40&$1.2\!\times\!10^5$&$1.7\!\times\!10^5$&31.89&7,237&\textbf{749}&\textbf{701}\\
Oort     &23.00&34.40&22.88&$1.4\!\times\!10^5$&$1.9\!\times\!10^5$&29.47&6,689&3,873&6,456\\
\textbf{PPO}&\textbf{19.64}&\textbf{29.92}&16.50&\textbf{$6.8\!\times\!10^4$}&\textbf{$8.1\!\times\!10^4$}&\textbf{10.59}&\textbf{2,404}&2,506&2,327\\
\bottomrule
\end{tabular*}
\end{table*}

\begin{table*}[t]
\caption{Image classification: 20 clients, 300 episodes; means across runs. Lower is better except accuracy.}
\label{tab:compact_image}
\centering
\scriptsize
\setlength{\tabcolsep}{5pt}
\begin{tabular}{lrrrrrr}
\toprule
Method & Accuracy & Latency/episode & Energy/episode & Uploaded/episode & Uploads & Downloads\\
\midrule
FedAsync&81.00\%&401.0 s&594.4 J&0.111 GB&39,602&1,227\\
FedBuff&80.81\%&\textbf{394.3 s}&\textbf{587.0 J}&0.111 GB&39,666&\textbf{768}\\
Oort&81.10\%&399.2 s&590.9 J&0.111 GB&39,572&1,334\\
DQN&62.09\%&436.3 s&599.2 J&\textbf{0.103 GB}&\textbf{36,504}&9,835\\
Always-sync&70.66\%&731.1 s&857.6 J&0.111 GB&39,610&39,319\\
\textbf{PPO}&\textbf{84.48\%}&406.3 s&595.7 J&0.111 GB&39,464&2,273\\
\bottomrule
\end{tabular}
\end{table*}

\begin{figure}[t]
\centering
\includegraphics[width=.8\linewidth]{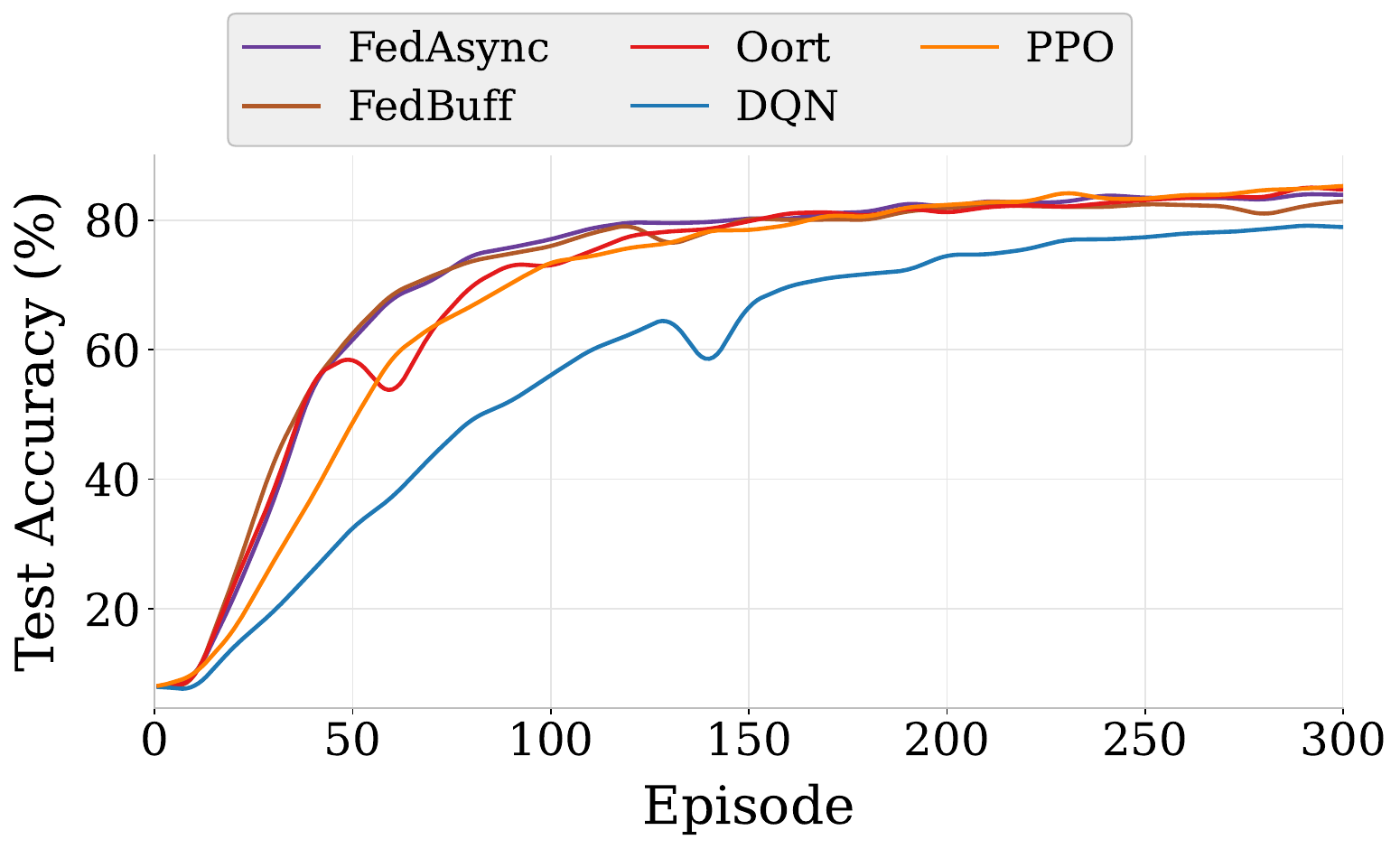}
\caption{Image-workload test-accuracy. PPO reaches the
highest final accuracy; DQN converges more slowly under the same 256-decision
rollout horizon.}
\label{fig:compact_accuracy}
\end{figure}

\subsection{Task Quality and System Cost}

Table~\ref{tab:compact_traffic} shows the clearest efficiency gain. PPO has the
lowest MAE and RMSE while using 2,404 uploads. Relative to the least expensive
non-PPO entry for each metric, it reduces latency by 43\%, energy by 52\%, and
uploaded traffic by 63\%. Its RMSE is 2.8\% below DQN, the strongest baseline
on that metric. DQN has better MAPE, illustrating that no policy
dominates every prediction measure. The principal result is therefore a joint
quality--cost improvement, not simply higher communication frequency.

Table~\ref{tab:compact_image} tests the same control interface on image classification workload. PPO reaches 84.48\% accuracy, a 4.17\% relative
improvement over Oort's 81.10\%, while latency and energy remain within
1.8\% and 0.8\% of Oort, respectively. FedBuff is the lowest-cost server
baseline; PPO's accuracy is 4.54\% higher relative to FedBuff's 80.81\%.
Always-sync incurs the most latency and energy and achieves 70.66\%
accuracy, 16.36\% lower relative to PPO. DQN uploads slightly fewer bytes
but achieves 62.09\% accuracy, 26.50\% lower relative to PPO.
Figure~\ref{fig:compact_accuracy} illustrates representative learning
trajectories; its endpoints differ from the means in Table~\ref{tab:compact_image}.

\subsection{Forecasting Backbone Validation}

\begin{figure}[t]
\centering
\includegraphics[width=\linewidth]{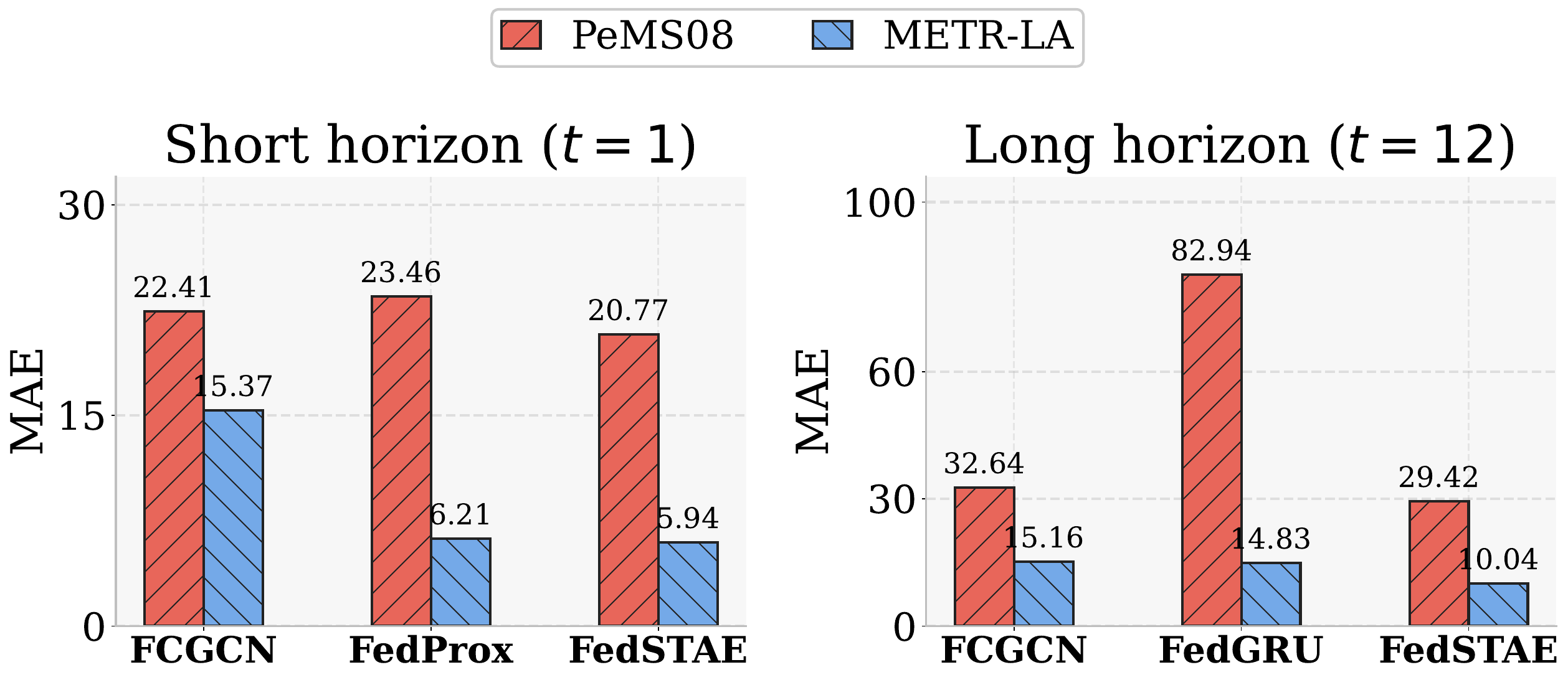}
\caption{FedSTAE forecasting quality against representative federated
baselines. Short and long horizons correspond to 5 and 60 minutes.}
\label{fig:compact_backbone}
\end{figure}

The synchronization result is meaningful only if its task model is a credible
federated forecasting backbone. We therefore compare FedSTAE with FCGCN,
FedProx, and FedGRU~\cite{liu2020privacy}. Figure~\ref{fig:compact_backbone}
reports one-step (5-minute) and 12-step (60-minute) MAE on PeMS-08 and
METR-LA~\cite{li2018dcrnn}, using a 60-minute input window. FedSTAE has the
lowest MAE among the shown methods on both datasets and horizons. The gain is
especially large at 12 steps on PeMS-08: 29.42 MAE versus 32.64 for FCGCN and
82.94 for FedGRU ~\cite{feng2024traffic,liu2024refol}. This check rules out the interpretation that communication
savings are measured around an intentionally weak traffic model.

Table~\ref{tab:compact_backbone_scale} isolates model quality as the PeMS-08
partition count grows. FedSTAE remains substantially better than FedGRU from
2 to 16 clients. Its long-horizon RMSE varies only from 29.21 to 30.60, although
MAE and MAPE gradually worsen as each client receives a smaller, more
heterogeneous portion of the sensor graph. This task-level scaling result
complements Table~\ref{tab:compact_secondary}(a), which measures
scheduler latency rather than forecast error.

\begin{table}[t]
\caption{Long-horizon PeMS-08 forecasting as client count increases.}
\label{tab:compact_backbone_scale}
\centering
\scriptsize
\setlength{\tabcolsep}{4pt}
\begin{tabular}{lrrrrrr}
\toprule
&\multicolumn{3}{c}{FedGRU}&\multicolumn{3}{c}{FedSTAE}\\
\cmidrule(lr){2-4}\cmidrule(lr){5-7}
Clients&MAE&RMSE&MAPE&MAE&RMSE&MAPE\\
\midrule
2  &58.58&78.26&61.54&\textbf{18.78}&\textbf{29.21}&\textbf{16.06}\\
4  &66.39&86.94&82.34&\textbf{19.10}&\textbf{29.71}&\textbf{15.79}\\
6  &61.39&82.94&75.23&\textbf{19.12}&\textbf{29.42}&\textbf{16.49}\\
8  &60.01&79.96&83.41&\textbf{19.93}&\textbf{30.60}&\textbf{17.28}\\
16 &56.86&75.31&76.01&\textbf{20.30}&\textbf{30.50}&\textbf{18.42}\\
\bottomrule
\end{tabular}
\end{table}

\subsection{Policy Ablations}

Table~\ref{tab:compact_ablation} compares policy configurations.
Accuracy is 11.72--14.65\% lower relative to Full (85.31\%).
The state-removal variants also disable bidirectional synchronization.
Relative to that reference (75.31\%), removing learning, privacy,
or synchronization state reduces accuracy by 2.07\%, 1.66\%,
and 3.32\%, respectively. Full also uses different reward and
optimizer settings, so these comparisons do not isolate the
effect of a single component.

\begin{table}[t]
\caption{Single-run image-policy ablations after 300 episodes.}
\label{tab:compact_ablation}
\centering
\scriptsize
\setlength{\tabcolsep}{3pt}
\begin{tabular}{lrrrr}
\toprule
Variant & Acc. & Latency & Energy & Uploaded\\
\midrule
Full & \textbf{85.31\%} &400.1 s&588.9 J&0.111 GB\\
w/o bidirectional &75.31\%&356.0 s&528.7 J&0.100 GB\\
w/o bidirectional + learning state &73.75\%&351.8 s&521.6 J&0.099 GB\\
w/o bidirectional + privacy state &74.06\%&356.8 s&529.7 J&0.100 GB\\
w/o bidirectional + sync state &72.81\%&355.3 s&527.0 J&0.100 GB\\
\bottomrule
\end{tabular}
\end{table}

\begin{figure}[t]
\centering
\includegraphics[width=.98\linewidth]{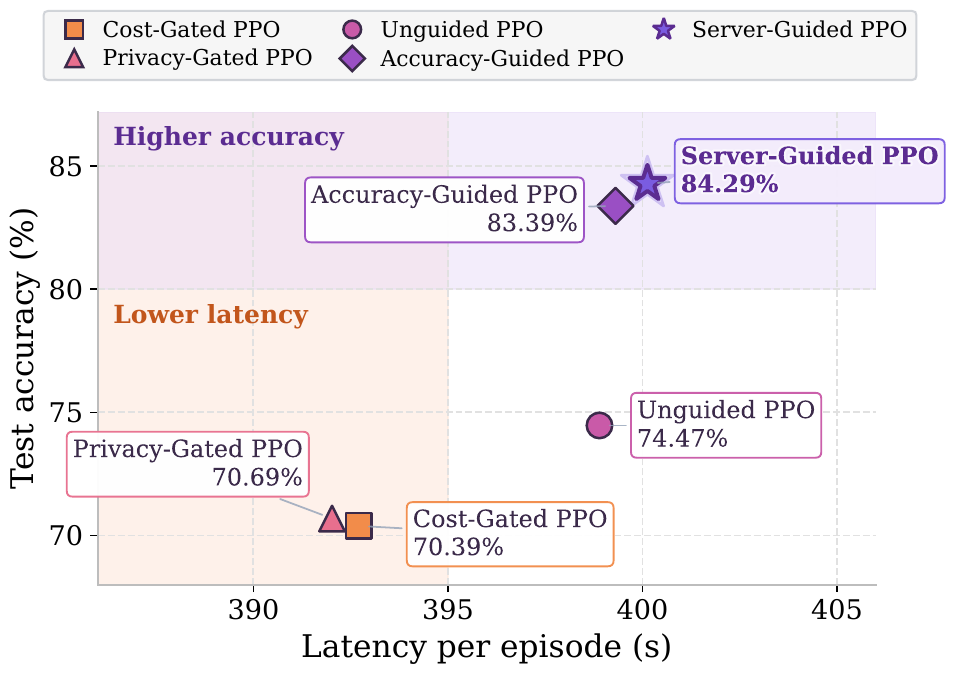}
\caption{PPO variant trade-off across 300-episode runs. Accuracy is reevaluated on the full test set using each run's validation-selected checkpoint; latency covers the full run. All five variants spend the same privacy budget \mbox{($\varepsilon=8.06$)}, so privacy is not encoded. Shading indicates lower latency and higher accuracy as visual guides.}
\label{fig:compact_frontier}
\end{figure}

\begin{table*}[t]
\caption{Scale and scheduler overhead. Scalability reports latency/episode (s).
Full-model sizes in (b) differ from transmitted payloads.}
\label{tab:compact_secondary}
\centering
\footnotesize
\begin{minipage}[t]{0.48\textwidth}
\centering
\textbf{(a) Clients and latency}\\[2pt]
\begin{tabular}{lrrrr}
\toprule
Method & 4 & 8 & 16 & 32\\
\midrule
FedAsync &375.6&384.6&378.8&390.3\\
FedBuff  &376.1&377.4&381.3&379.9\\
Oort     &377.8&384.9&379.5&396.7\\
Cost-Gated PPO&\textbf{362.9}&\textbf{365.3}&\textbf{363.2}&\textbf{365.7}\\
\bottomrule
\end{tabular}
\end{minipage}\hfill
\begin{minipage}[t]{0.48\textwidth}
\centering
\textbf{(b) Deployment overhead}\\[2pt]
\begin{tabular}{lrrr}
\toprule
Component & Params & MB & Runtime\\
\midrule
PPO &20.4k&0.081&0.194 ms\\
STAEFormer&1.22M&9.79&18.85 ms\\
ResNet-18&11.17M&44.70&33.67 ms\\
\bottomrule
\end{tabular}
\end{minipage}
\end{table*}

\subsection{Privacy, Scale, and Overhead}

Figure~\ref{fig:compact_frontier} separates five policies with the same action
interface. Unguided PPO uses only client observations; Accuracy-Guided and
Server-Guided PPO add client priors and compact server broadcasts. Cost-Gated
and Privacy-Gated PPO increase communication and privacy pressure. We use
Server-Guided PPO for the accuracy-oriented image comparison and Cost-Gated PPO
for the privacy and scale studies. Because a scalar reward selects only one
multi-objective operating point, every reported configuration is chosen on
held-out validation data rather than treated as universally best.

The gated configurations reduce latency by approximately
1.6--1.7\% relative to Unguided PPO, at lower accuracy.
Server-Guided PPO achieves the highest validation-selected
test accuracy (84.29\%), compared with Unguided PPO's 74.47\%.
These results compare complete policy configurations rather than
isolating the effect of guidance alone. Privacy expenditure is identical across the five PPO variants (\(\varepsilon = 8.06\)),
and DQN spends \(\varepsilon = 7.73\).

The client-count sweep in Table~\ref{tab:compact_secondary}(a) increases the
federation from 4 to 32 clients. Cost-Gated PPO remains the lowest-latency
method at every point; its latency changes by only 0.8\% from 4 to 32 clients,
whereas the compared baselines are both higher and more variable. This supports
the intended parameter sharing: additional clients do not enlarge the policy.
Finally, Table~\ref{tab:compact_secondary}(b) shows that one policy decision
takes 0.194 ms and 0.081 MB, much smaller than either task model. Scheduler
inference is therefore negligible relative to local prediction and network
transfer in this environment.

\section{Discussion}

The experiments distinguish client-side synchronization from server-side
aggregation. We leave Eq.~\eqref{eq:fedavg_overview} unchanged and obtain the
reported trade-offs solely by changing whether a client uploads, downloads,
does both, or waits. The matched gated and unguided runs show a modest latency–accuracy trade-off, with no reduction in aggregate-release privacy expenditure. Small server broadcasts provide context without moving the action back to the server.

\subsection{Why Separate Upload and Download?}

The action counts help explain the benefit of separating communication
directions. In traffic forecasting, PPO uses 2,404 uploads and 2,506 downloads;
it neither collapses to Always-sync nor treats every server refresh as requiring
a local contribution. FedBuff performs only 749 downloads and 701 aggregations,
but has 9.2\% higher RMSE and nearly three times as many uploads as PPO. This
comparison separates two forms of sparsity: reducing server aggregation alone
does not ensure that clients send only useful updates, and reducing downloads
too aggressively can leave local models behind the global state. The four-way
action space lets the policy manage these effects independently.

The image results expose a different operating regime. PPO's upload count is
close to the server-side baselines, but it uses far fewer downloads than
Always-sync and reaches higher accuracy. Here, the value lies less in suppressing
all communication than in choosing the direction and timing of synchronization.
This difference across tasks is exactly why a fixed period is brittle: the
traffic workload rewards a strongly cost-gated pattern, whereas the non-IID
image workload benefits from a more accuracy-oriented schedule.

\subsection{Control Placement and Deployment}

The scheduler is client-executed but not client-isolated. A server may broadcast
small coordination features with its model version, after which the client
combines that context with signals unavailable at the server, such as current
loss change, dirty progress, channel cost, and remaining privacy budget. The
server does not issue the action. This division preserves local responsiveness
while avoiding the assumption that a client can infer federation-wide freshness
from local history alone.

A deployment would first select normalized cost weights from service-level
goals, for example an energy ceiling or a target privacy budget, and train the
shared policy against representative availability and channel traces. At
runtime, each client constructs the observation only when idle, applies the
hard feasibility mask, and executes one policy forward pass. The server needs
no policy-specific aggregation rule. If resource distributions drift outside
the training range, the safe fallback is to retain masking and use a conservative
schedule while the policy is revalidated; the current experiments do not claim
zero-shot robustness to arbitrary drift.

\subsection{Scope of Generalization}

The traffic and image experiments reuse the synchronization abstraction,
observation groups, legal-action logic, and policy architecture. They do not
reuse one set of learned weights across tasks. Each policy is trained for its
workload because the meaning and scale of task utility differ between
forecasting error and classification loss. The evidence therefore supports
\emph{mechanism-level} portability, not cross-workload policy transfer. This is
still operationally useful: adding a task model does not require designing a
new synchronization protocol, but it does require training and validating the
controller for that task's dynamics.

\section{Limitations}

Our evaluation is an emulated edge-federated setting rather
than a live deployment. First, communication, availability, and energy
are evaluated in an event-driven environment with calibrated analytical costs,
not a geographically distributed device deployment. Second, privacy claims are conditional on the fixed clipping, aggregate-noise mechanism, accounting scope, and budget above; PPO schedules contributions but does not strengthen the underlying privacy guarantee. Third, the scale study stops at 32 clients. Trace-driven tests on
physical edge devices, more seeds, and larger federations are necessary before
making production-level latency, energy, or robustness claims.


\section{Conclusion}

Client-side synchronization turns communication from a fixed protocol choice
into a resource-aware decision. Across traffic forecasting and image
classification, the shared masked-PPO scheduler improves the measured
quality--cost trade-off while retaining a fixed aggregation rule. It also incorporates explicit privacy-budget constraints and adds little local runtime or memory overhead.

\clearpage

\bibliographystyle{IEEEtran}
\bibliography{main}

\end{document}